\documentclass[10pt,aps,prb,raggedbottom,longbibliography,reprint,citeautoscript,letterpaper,superscriptaddress]{revtex4-2}

\usepackage{soul}

\usepackage{graphicx}
\usepackage{dcolumn}
\usepackage{bm}
\usepackage{xr}
\usepackage[usenames,dvipsnames]{color}
\usepackage{microtype}
\usepackage{multirow}
\usepackage{longtable}
\usepackage{lmodern}
\usepackage[bookmarks=false,colorlinks]{hyperref}
\hypersetup{
    linkcolor=magenta,        
    citecolor=Plum,     
    filecolor=Plum,      	
    urlcolor=MidnightBlue,           
}

\newcommand{\sfo}{SrFeO$_3$}
\newcommand{\sfox}{SrFeO$_x$}

\newcommand{\sfobm}{SrFeO$_{2.5}$}
\newcommand{\sto}{SrTiO$_3$}
\newcommand{\lao}{LaAlO$_3$}

\newcommand{\pds}{$pd\sigma$}

\begin{document}

\preprint{APS/123-QED}

\title{Biaxial Strain Control of Helimagnetism via Chemical Expansion in Thin Film \sfo}
\author{J. Fowlie}
  \email{jfowlie@northwestern.edu.}

\affiliation{ 
Stanford Institute for Materials and Energy Sciences, SLAC National Accelerator Laboratory, Menlo Park, CA 94025, USA%
}%
\affiliation{ 
Department of Applied Physics, Stanford University, Stanford, CA 94305, USA%
}%
\affiliation{ 
Department of Materials Science and Engineering, Northwestern University, Evanston, IL 60208, USA%
}%

\author{J. Li}
\affiliation{ 
Stanford Institute for Materials and Energy Sciences, SLAC National Accelerator Laboratory, Menlo Park, CA 94025, USA%
}%
\affiliation{ 
Department of Applied Physics, Stanford University, Stanford, CA 94305, USA%
}%

\author{D. Puggioni}
\affiliation{ 
Department of Materials Science and Engineering, Northwestern University, Evanston, IL 60208, USA%
}%

\author{L. Barreto}
\affiliation{Singh Center for Nanotechnology, University of Pennsylvania, Philadelphia, PA 19104, USA}
\affiliation{Center for Natural Sciences and Humanities, Federal University of ABC - UFABC, Santo André 09210-580, SP, Brazil}
\affiliation{Department of Physics, University of Johannesburg, PO Box 524, Auckland Park, 2006, South Africa}

\author{L. Yuan}
\affiliation{ 
Department of Materials Science and Engineering, Northwestern University, Evanston, IL 60208, USA%
}%

\author{J. M. Rondinelli}
\affiliation{ 
Department of Materials Science and Engineering, Northwestern University, Evanston, IL 60208, USA%
}%

\author{R. Sutarto}
\affiliation{ 
Canadian Light Source, Saskatoon, SK S7N 2V3, Canada%
}%

\author{T. D. Boyko}
\affiliation{ 
Canadian Light Source, Saskatoon, SK S7N 2V3, Canada%
}%

\author{F. Orlandi}
\affiliation{ 
ISIS Neutron and Muon Source, STFC Rutherford Appleton Laboratory, Harwell Campus, Didcot, Oxfordshire OX11 0QX, United Kingdom%
}%

\author{P. Manuel}
\affiliation{ 
ISIS Neutron and Muon Source, STFC Rutherford Appleton Laboratory, Harwell Campus, Didcot, Oxfordshire OX11 0QX, United Kingdom%
}%

\author{D. D. Khalyavin}
\affiliation{ 
ISIS Neutron and Muon Source, STFC Rutherford Appleton Laboratory, Harwell Campus, Didcot, Oxfordshire OX11 0QX, United Kingdom%
}%

\author{E. G. Lomeli}
\affiliation{ 
Stanford Institute for Materials and Energy Sciences, SLAC National Accelerator Laboratory, Menlo Park, CA 94025, USA%
}%
\affiliation{Department of Materials Science and Engineering, Stanford University, Stanford, CA 94305, USA%
}%

\author{B. Moritz}
\affiliation{ 
Stanford Institute for Materials and Energy Sciences, SLAC National Accelerator Laboratory, Menlo Park, CA 94025, USA%
}%

\author{T. P. Devereaux}
\affiliation{ 
Stanford Institute for Materials and Energy Sciences, SLAC National Accelerator Laboratory, Menlo Park, CA 94025, USA%
}%
\affiliation{Department of Materials Science and Engineering, Stanford University, Stanford, CA 94305, USA%
}%

\author{S. Koroluk}
\affiliation{Department of Physics \& Engineering Physics, University of Saskatchewan, Saskatoon, Canada S7N 5E2}

\author{R. J. Green}
\affiliation{Department of Physics \& Engineering Physics, University of Saskatchewan, Saskatoon, Canada S7N 5E2}
\affiliation{Stewart Blusson Quantum Matter Institute, University of British Columbia, Vancouver, Canada V6T 1Z4}

\author{S. J. May}
\affiliation{Department of Materials Science and Engineering, Drexel University, Philadelphia, Pennsylvania 19104, USA}

\author{H. Y. Hwang}
\email{hyhwang@stanford.edu.}
\affiliation{ 
Stanford Institute for Materials and Energy Sciences, SLAC National Accelerator Laboratory, Menlo Park, CA 94025, USA
}%
\affiliation{ 
Department of Applied Physics, Stanford University, Stanford, CA 94305, USA%
}%

\date{\today}

\begin{abstract}

We demonstrate control of helimagnetic order in biaxially strained \sfo{} thin films using neutron diffraction and resonant soft x‑ray scattering. \sfo{}, a negative charge‑transfer oxide, exhibits a complex magnetic phase diagram that includes multi‑$q$ spin structures. Tensile epitaxial strain produces a pronounced shortening of the helimagnetic ordering length and a tilting of the magnetic ordering vector. We interpret this behavior in terms of chemical expansion: lattice dilation under tensile strain lowers the energetic cost of oxygen vacancies, leading to an expanded unit cell that modifies Fe--O hybridization and enhances superexchange relative to double exchange. These results reveal how epitaxial strain can indirectly tune helimagnetism through defect-driven chemical expansion, highlighting the strong coupling between lattice, chemistry, and magnetic order in transition‑metal oxides. Our findings establish chemical expansion as an effective mechanism for engineering complex magnetic textures in oxide thin films, with implications for spintronic, magnonic, and quantum information applications.

\end{abstract}

\maketitle

\section{\label{sec:level1}Introduction}

Helimagnets are a class of magnetically ordered materials in which local moments arrange in a spiral or a helical pattern. Despite having no \textit{net} magenetization, these systems can potentially host chiral spin textures, and, in multi-q helimagnets, topologically-protected magnetic structures.
As a result, helimagnets are gaining interest for potential spintronic and magnonic components \cite{Islam2023} as well as future qubit platforms \cite{Psaroudaki2023}. They may also exhibit non-relativistic spin-momentum coupling with p-wave symmetry \cite{Hayami2022,Hodt2025}.

Helimagnetism typically arises in magnetic systems where the crystalline structure breaks inversion symmetry. This is the case in B20 structures, where a Dzyaloshinskii-Moriya interaction (DMI) emerges and the helimagnetic ordering length can be tens of nanometers \cite{Muhlbauer2009,Yu2010}. Helimagnetism can also occur in centrosymmetric crystals, i.e., without the magnetic chirality being locked to the underlying crystal chirality. In these cases, the helimagnetism may stem from a geometric frustration of the magnetic order and the ordering length is typically much shorter, only a few nanometers \cite{Kurumaji2019}.
\sfo{} is an example of a helimagnetic compound that is centrosymmetric and cubic, so its helimagnetism cannot be explained by either DMI physics or geometric frustration \cite{Takeda1972}, but, rather, may arise from the competition between exchange interactions \cite{Mostovoy2005}. The higher symmetry of the chemical unit cell, as well as the metallicity of \sfo{}, make it an attractive material for a future reconfigurable magnetic component.

The helimagnetic order in \sfo{} is aligned approximately along the [1 1 1] direction and has an incommensurate length of 17--18 Å. \sfo{} is reported to have multiple helimagnetic phases as a function of temperature \cite{Ishiwata2011,Chakraverty2013,Onose2020}. Ishiwata and coworkers recently proposed that \sfo{} hosts a double-q state (proper screw and cycloid) at low temperature and a quadruple-q state (4 $\times$ proper screw) for 80 K $\lesssim$ T $\lesssim$ 120 K, the latter of which is speculated to host a topologically-protected three-dimensional skyrmion lattice \cite{Ishiwata2020}. The characteristic length scale of these magnetic textures is expected to be on the order of 5 nm, while the helimagnetic pitch itself is shorter than 2 nm \cite{Ishiwata2020}. These exceptionally short length scales make direct real-space imaging of the magnetic order in \sfo{} particularly challenging.

The leading theory for the origin of helimagnetism in \sfo{} is based on characteristics of its electronic structure \cite{Mostovoy2005}. Nominally, \sfo{} is high spin $d^4$, and calculated to be nearly half-metallic \cite{Rondinelli2010b}. Experimentally, \sfo{} exhibits significant charge transfer character \cite{Green2024}. The transfer of electrons from oxygen to iron make it closer to $d^5\underline{L}$ \cite{Rogge2019}, where $\underline{L}$ denotes an oxygen ligand hole. X-ray spectroscopy measurements suggest a $d^5\underline{L}$ portion as high as 90\% \cite{Rogge2019b} while recent multiplet calculations reveal greater than 60\% of the $d^5\underline{L}$ character is Fe high spin $S = 5/2$ configuration with an antiparallel oxygen $S = 1/2$ \cite{Takegami2024}. It has been suggested that, in this negative charge transfer energy regime, the interaction between localized electronic moments and itinerant oxygen holes leads to the preferential helimagnetic order \cite{Mostovoy2005,Li2012a}.

Going from bulk to thin film samples introduces a symmetry lowering and biaxial strain, which may influence the electronic structure, and therefore the helimagnetism. 
From existing literature, helimagnetic, spiral magnet, chiral magnet, or skyrmion lattice compounds respond to the thin film environment in complex and distinct ways depending on the underlying physics. In conventional DMI-mediated B20 materials, imposing thin film geometry is reported to extend the temperature and field parameter space over which the skyrmion lattice is observed \cite{Huang2012,Seki2017,Budhathoki2020}. This is suggested to be due to induced magnetic anisotropy and is supported by uniaxial pressure studies on single crystals \cite{Chacon2015}. In B20 compounds, hydrostatic pressure has also been observed to cause a shortening of the chiral magnetic ordering length \cite{Fak2005} and a reorientation of the ordering vector, attributed to enhanced electron itinerancy \cite{Bannenberg2019}. In multiferroic spiral magnets, the strong magnetoelastic coupling provides a direct path for strain-induced modifications to the magnetic order. For instance, in BiFeO$_3$ the spin cycloid is suppressed in favor of a collinear magnetic order when biaxial strain is high and either tensile or compressive \cite{Sando2013}. In cobalt-doped BiFeO$_3$ thin films, the spiral ordering length is also observed to increase significantly with decreasing thickness \cite{Burns2019}. In TbMnO$_3$, the incommensurate cycloid shortens and becomes commensurate under compressive biaxial strain \cite{Shimamoto2017}, while in CaMn$_7$O$_{12}$, the incommensurate spiral retains the same length in thin films as in bulk \cite{Huon2018}.

The physics underlying the helimagnetism of \sfo{} is expected to be distinct from either B20 or multiferroic compounds because \sfo{} is centrosymmetric and thus, to leading order, lacks DMI and significant magnetoelastic coupling. Exploration of the effect of the thin film environment on a centrosymmetric helimagnet has, to our knowledge, not been reported.
Here we show, using neutron and X-ray scattering, that the helimagnetic ordering is strongly affected by the biaxial strain imposed by the substrate in \sfo{} thin films. Specifically, we first report that the thin film environment preferentially selects a magnetic domain where the helimagnetic ordering vector tilts towards the more compressed direction. We then show that the helimagnetic real-space ordering length decreases as the unit cell volume increases, in contrast to the effects of strain or pressure on B20 or multiferroic helimagnetic compounds. Such a shortening may be explained by a relative enhancement of the superexchange interaction over the double exchange interaction. Our density functional theory (DFT) calculations show that the shortening is unlikely to be caused by the direct strain effect and instead may stem from a strain-induced shift in the chemical potential driven by oxygen stoichiometry.
In order to demonstrate the robustness of our findings across different samples, we study \sfo{} thin films grown and annealed in different oxidising environments: \sfo{} grown by pulsed laser deposition (PLD) and post-annealed in ozone and \sfo{} grown by molecular beam epitaxy (MBE) and post annealed in oxygen plasma.

\section{Experiment}

\subsection{Sample preparation}

\begin{figure}
\includegraphics{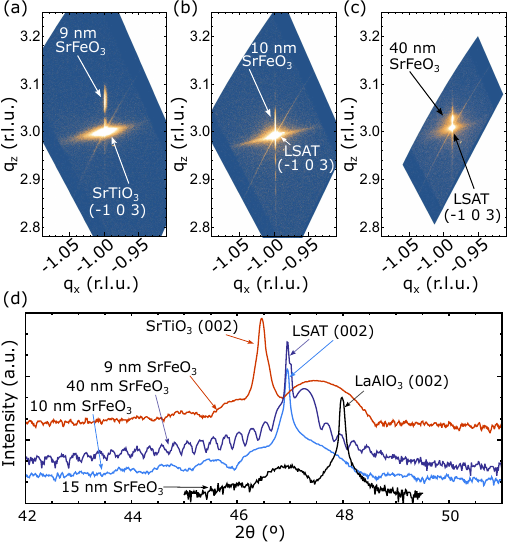}
\caption{\label{xrd}a)-c) Reciprocal space maps around the (-1 0 3) peak of the substrate: (a) 9 nm PLD-grown \sfo{}/\sto{}, (b) 10 nm PLD-grown \sfo{}/LSAT and (c) 40 nm PLD-grown \sfo{}/LSAT. d) (0 0 2) peak $\theta$-2$\theta$ measurements of 15 nm MBE-grown \sfo/\lao{} (black), 10 nm PLD-grown \sfo{}/LSAT (light blue) and 9 nm PLD-grown \sfo{}/\sto{} (red) used for synchrotron x-ray measurements and 40 nm PLD-grown \sfo{}/LSAT used for neutron scattering (dark blue).}
\end{figure}

Thin film samples were grown by PLD or MBE and post-processed according to the procedure outlined in Methods. Substrates are (001)-oriented \sto{}, (001)-oriented (LaAlO$_3$)$_{0.3}$(Sr$_2$AlTaO$_6$)$_{0.7}$ (LSAT) and (001)-oriented \lao{}. For synchrotron soft x-ray scattering we prepared: 9 nm \sfo{}/\sto{} and 10 nm \sfo{}/LSAT grown by PLD and 15 nm \sfo{}/\lao{}, 15 nm \sfo{}/LSAT, and 14 nm \sfo{}/\sto{} grown by MBE. We also prepared eight equivalent 40 nm \sfo{}/LSAT samples by PLD for neutron diffraction.

Fig. \ref{xrd} shows characterization by in-house x-ray diffraction reciprocal space maps (RSMs) and $\theta$-2$\theta$ measurements. RSMs indicate that the films are fully epitaxially-strained. $\theta$-2$\theta$ scans show finite thickness oscillations, suggesting a high crystalline quality of the epitaxial \sfo{}. Table \ref{paramtab} summarizes relevant structural parameters for the samples discussed in this work, as well as bulk samples reported previously \cite{Takeda1972}. The c-axis (out-of-plane, or along the surface normal) lattice parameters have been obtained from the $\theta$-2$\theta$ scans.

\begin{table*}
\begin{ruledtabular}
\begin{tabular}{rcccccc}
 &Bulk&\lao{} (MBE)&LSAT (MBE) &\sto{} (MBE)&LSAT (PLD)&\sto{}(PLD)\\
\hline
a,b-parameter &3.85 Å&3.79 Å&3.87 Å&3.905 Å&3.87 Å&3.905 Å\\
c-parameter &3.85 Å&3.876 Å&3.831 ÅX&3.81 Å&3.84 Å&3.81 Å\\
Epitaxial strain & 0\% &-1.6\%&+0.4\%&+1.4\%&+0.4\% & +1.4\%\\
Unit cell volume &57.07 Å$^3$&55.68Å$^3$&57.37 Å$^3$&58.10 Å$^3$&57.57 Å$^3$&58.10 Å$^3$\\
$[1 1 1]$ diagonal length &6.67 Å&6.61 Å&6.68 Å&6.71 Å&6.69 Å&6.71 Å\\
\end{tabular}
\end{ruledtabular}
\caption{\label{paramtab}Unit cell a, b, and c lattice parameters, epitaxial strain, unit cell volume, and [1 1 1] body diagonal length for \sfo{} in bulk \cite{Takeda1972}, and as a thin film grown on \lao{}, LSAT, and \sto{} by PLD and by MBE. All values are reported at room temperature.}
\end{table*}

\begin{figure}
\includegraphics{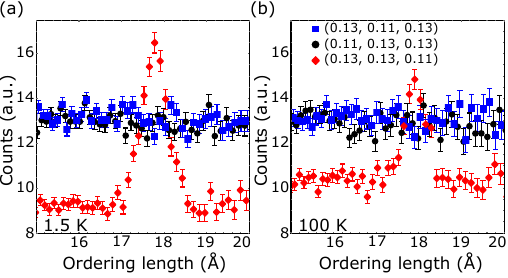}
\caption{\label{neutron}$d$-spacing neutron diffraction data around the magnetic Bragg peak of \sfo{}. (a) At 1.5 K and (b) at 100 K. Both panels show integrated intensity within three regions of reciprocal space that are equivalent in cubic symmetry; (0.13, 0.13, 0.11), (0.11 0.13 0.13) and (0.13 0.11 0.13).}
\end{figure}

\subsection{Neutron diffraction}
In order to achieve a sufficient sample volume for neutron scattering, eight \sfo{} samples were prepared (see Methods) on 5 $\times$ 5 mm LSAT substrates, with a film thickness of 40 nm giving a total sample volume of 8 $\times$ 10$^{-3}$ mm$^3$. The samples were then mounted in a mosaic, aligned relative to each other within an azimuthal angle of 1º, for neutron experiments carried out at the WISH diffractometer at the ISIS Neutron Source \cite{Chapon2011}.

\begin{figure*}
\includegraphics{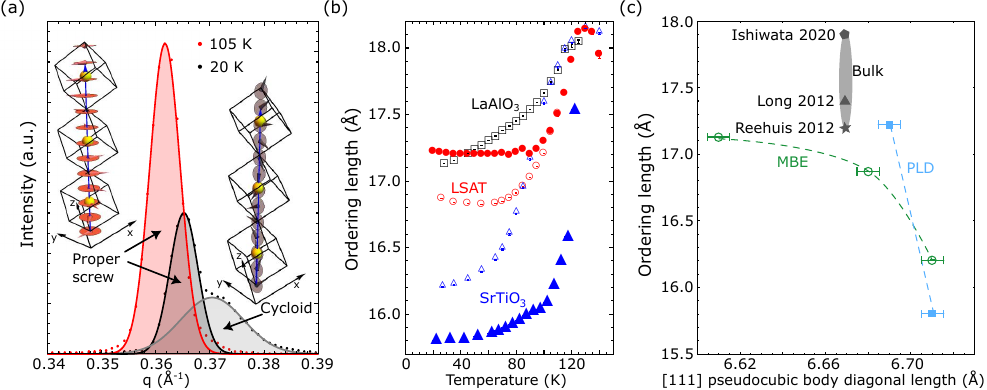}
\caption{\label{rsxs}Resonant soft x-ray scattering data recorded on the Fe L$_3$ edge in linear horizontal ($\pi$) polarization. (a) $\theta$-2$\theta$ scans across the magnetic Bragg peak of PLD-grown \sfo{}/LSAT. 20 K (black) and 105 K (red) data are shown. The more intense peak corresponds to the proper screw order, while the less intense peak corresponds to a spin cycloid. Solid lines and shaded regions are for illustrative purposes and are not numerical fits. Insets sketch the different spiral structures where the incommensurate ordering vector is shown in blue and iron atoms in the body center positions are indicated by yellow spheres. (b) The proper screw ordering length extracted from the $\theta$-2$\theta$ scans as a function of temperature. PLD-grown samples are represented by filled shapes while MBE-grown samples are represented by empty shapes. Error bars (often smaller than the plot marker) represent one standard deviation from the center of a Gaussian lineshape. (c) The base temperature ($\approx$ 20 K) ordering length of the proper screw structure as a function of the [111] body diagonal length of the chemical unit cell for the two series of sample. A survey of bulk values is represented by the gray plot markers \cite{Reehuis2012,Long2012,Ishiwata2020}. For consistency, body diagonal length is calculated from room temperature lattice parameters -- \sfo{} exhibits only a weak thermal contraction of $\approx -0.25 \%$ from room temperature to 100 K \cite{Takeda1972}. Horizontal error bars represent a $\pm$ 0.005 Å tolerance on the unit cell diagonal while vertical error bars are smaller than the plot markers. The dashed lines are guides to the eye.
}
\end{figure*}

Fig. \ref{neutron} shows the diffraction focused neutron data as a function of d-spacing at the position (0,0,0)+k, where k = (0.13,0.13,0.11), (0.13,0.11,0.13), (0.11,0.13,0.13). Panel a) is data recorded at 1.5 K and panel b) is data recorded at 100 K. The ordering length of $\approx$ 17.8 Å is consistent with previous reports. In both panels, three curves are shown, corresponding to three regions of reciprocal space where equivalent peaks, related by threefold symmetry, would be expected in a cubic structure. The Miller indices corresponding to the Bragg peaks were calculated absolutely from an orientation matrix based on the LSAT structural peaks that were visible in the same detector image. The same peak is observed on a single sample rather than a mosaic, as shown in Fig. S1 in the Supplemental Materials (SM). The same indexation of the magnetic peaks has been obtained in both cases confirming that the mosaic alignment spread does not affect the determination of the propagation vector.

The magnetic Bragg peak observed is (0.13 0.13 0.11). The observation that the ordering is tilted slightly away from the [1 1 1] direction was previously reported by Ishiwata and co-workers in the low temperature, double-q state, who observed three peaks of the form ($q$, $q$, $q'$), ($q$, $q'$, $q$), and ($q'$, $q$, $q$) where $q' > q$ \cite{Ishiwata2020}. Compared to those bulk reports we see three differences: 1) the peak we observe is of the form ($q$, $q$, $q'$) but with $q > q'$. 2) We do not observe the peaks ($q$, $q'$, $q$) and ($q'$, $q$, $q$). 3) The same authors report that above 80 -- 90 K the ordering is uniquely along the high symmetry [1 1 1] direction. This is not the case in our experiment and the peak remains at (0.13, 0.13, 0.11).

Considering magnetic symmetry, even with these three differences, our observations are not inconsistent with the symmetries determined by Ishiwata and coworkers. Our low temperature (Phase I) magnetic structure is consistent with at least two domains of double-$q$ ordering. Ishiwata and coworkers interpreted their Phase II magnetic structure as a single domain of a quadruple-$q$ order. Our data is consistent with this, but in their case the propagation vector is along the high symmetry [1 1 1] direction so the magnetic structure retains cubic symmetry while our Phase II magnetic structure must have, at most, tetragonal symmetry.

The (0.13, 0.13, 0.11) peak that we observe corresponds to a tilting of the magnetic ordering vector away from the [1 1 1] direction more out of the plane of the sample. Due to the tensile strain, the out-of-plane direction is more compressed. Although biaxial strain breaks the symmetry between the in-plane and out-of-plane directions, it is notable that the system so strongly prefers only one of the three ordering directions since the strain is moderate and the films are relatively thick. This suggests a high sensitivity of the helimagnetic order to the biaxial strain.

\subsection{Resonant soft x-ray scattering}

Resonant soft x-ray scattering (RSXS) measurements were carried out at the REIXS beamline of the Canadian Light Source \cite{Hawthorn2011}. The samples were mounted on a wedge to bring the [111] direction into the diffraction plane. All measurements were $\theta$-2$\theta$ scans around the position of the magnetic Bragg peak conducted with a beam energy resonant with the Fe L$_3$ edge at around 708 eV. Fig. \ref{rsxs}a  presents a subset of the data at 20 K and at 100 K for PLD-grown \sfo{}/LSAT. The full set of measurements for all samples and temperatures are presented in Fig. S2 in the SM. At low temperatures there are two peaks observed for some samples. It is deduced from linear dichroism (see SM Fig. S3) that the peak at higher momentum transfer, $q$, (shorter ordering length) is a cycloidal spiral while the more intense peak at lower $q$ (longer ordering length) is a proper screw. We note that our neutron scattering experiment does not resolve the separation between the two peaks but, as shown in SM Fig. S1, the width is consistent with the two close peaks observed in RSXS. At intermediate temperatures there is only one peak clearly observed in all samples, deduced to be of proper screw structure.
The data obtained is consistent with Ishiwata and coworkers' neutron diffraction reports on bulk single crystal samples, but to our knowledge, this is the first confirmation of an equivalent multi-q ground state in thin film samples.

Fig. \ref{rsxs}b plots the ordering length as a function of temperature for the full data set. Only the peak arising from the proper screw structure is plotted for simplicity but both peaks have the same trend with temperature. The ordering length increases with increasing temperature for all \sfo{} samples, until, around 120 K, the magnetic Bragg peak is extinguished and \sfo{} becomes paramagnetic.
Finally, Fig. \ref{rsxs}c shows the proper screw ordering length as a function of the pseudocubic unit cell body diagonal as stated in Table \ref{paramtab}. In the samples with two peaks, the cycloidal spiral has the same trend with strain as the proper screw, as can be seen in SM Fig. S2. The proper screw ordering length decreases on the order of 10\% when the unit cell is expanded by tensile biaxial strain by approximately only 1\%. This is an extreme strain effect.

\section{Theory}

From both our neutron diffraction and our resonant soft x-ray diffraction results we see that the incommensurate helimagnetic spiral of \sfo{} extends when the lattice is more compressed, and contracts significantly when tensile strain causes the unit cell to expand. We now discuss the possible origins with support from first principles DFT (see Methods).

In the extreme, long wavelength helimagnetic order (small helimagnetic angle) begins to resemble ferromagnetic order, while short helimagnetic order begins to resemble antiferromagnetic order (collinear from a spin cycloid and non-collinear from a proper screw). In these terms, the magnetic ordering in \sfo{} moves closer to an antiferromagnetic ground state when the lattice is expanded and closer to a ferromagnetic ground state when the lattice is compressed. This is consistent with M\"{o}ssbauer spectroscopy results showing that under a hydrostatic pressure of 13 GPa, where the lattice is more compressed, \sfo{} transitions to a fully ferromagnetic ground state \cite{Nasu1993}. According to the model developed by Mostovoy \cite{Mostovoy2005}, helimagnetic order arises from a competition between the kinetic energy of itinerant, hybridized Fe–O $p-d$ bands and the covalent energy gain associated with 
$p-d$ hybridization. Although this mechanism is formulated in terms of band energetics rather than localized exchange interactions, its magnetic consequences are commonly interpreted in terms of an effective competition between double-exchange (DE)–like (kinetic) and superexchange (SE)-like (covalent) tendencies \cite{Li2012a, Li2012b}. In this effective picture, the DE-like contribution favors long-wavelength (ferromagnetic-like) order, whereas the SE-like contribution favors shorter-wavelength (antiferromagnetic-like) order.

\begin{figure}
    \centering
\includegraphics[width=\columnwidth]{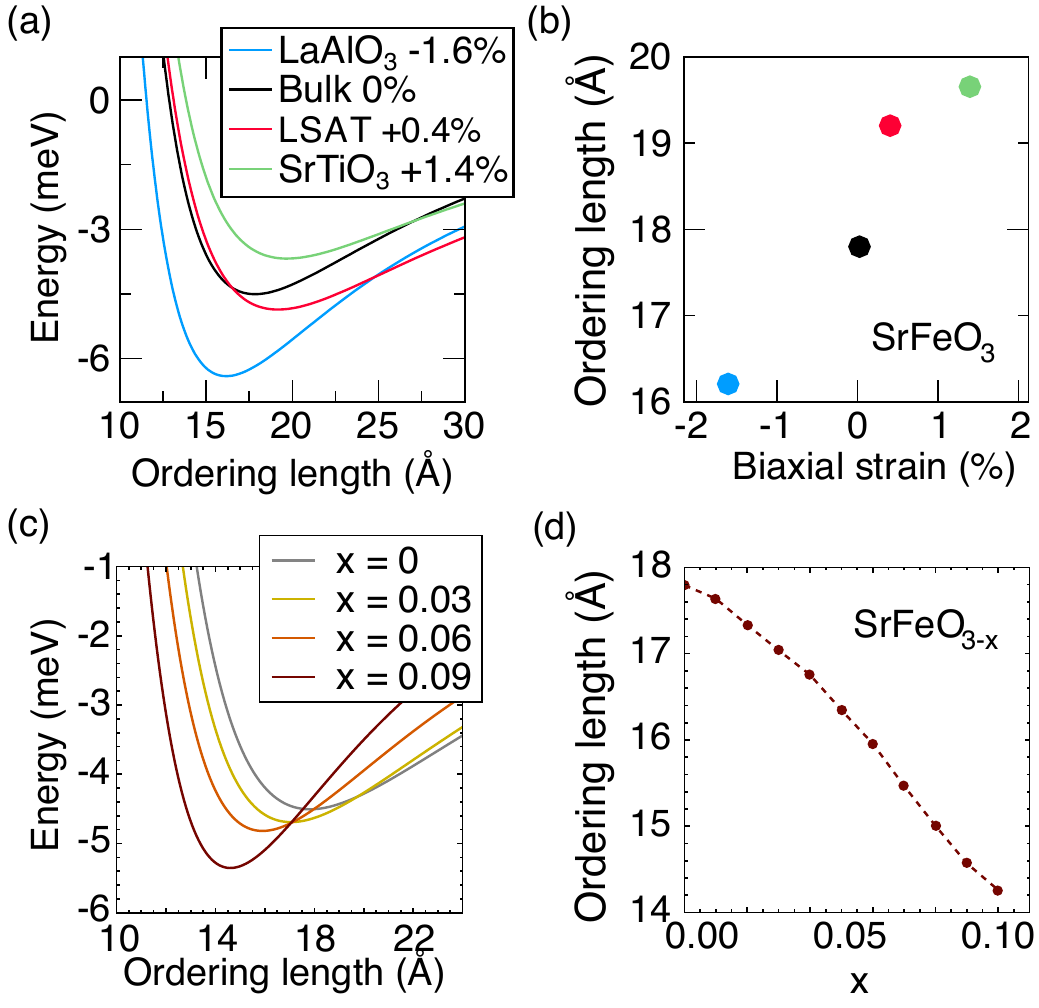}
    \caption{Strain dependence of the [111] helimagnetic ordering length from DFT for (a,b) pristine SrFeO$_3$ under different strains and (c,d) oxygen deficient bulk SrFeO$_{3-x}$.(a, c) display the total energy difference between the helimagnetic state (propagating along [111]) and the ferromagnetic state while (b, d) display the ordering length as a function of strain and $x$ respectively.}
    \label{fig:strain+Ov}
\end{figure}

We consider the electronic structure of pristine \sfo{} under strain. Fig. \ref{fig:strain+Ov}a shows the ordering length dependence of the total energy difference between the helimagnetic ordering along the [111] direction and the ferromagnetic state, evaluated for various epitaxial strains. Across the full strain range, from -1.6\% compressive (\lao{}) to +1.4\% tensile (\sto{}), the helical [111] configuration is consistently lower in energy than the ferromagnetic state, indicating that it is generally favorable. The position of the energy minimum determines the helimagnetic ordering length (Fig. \ref{fig:strain+Ov}b), and indicates that it \textit{increases} with tensile strain. This result can be understood by considering the direct effect of strain on the electronic structure. Briefly, 
tensile strain on negative charge transfer \sfo{} reduces the orbital overlap between the iron 3$d$ $e_g$ states and oxygen $p$ states, quantified by the integral \pds{}. DE scales approximately as $\frac{(pd\sigma)^2}{\Delta}$, where $\Delta$ is the charge-transfer energy. SE scales as $\frac{(pd\sigma)^4}{U \Delta^2}$, where $U$ is the on-site Coulomb repulsion of the Fe $3d$ states~\cite{Slater1954,Goodenough:1955,Kanamori:1959,Anderson1950,Mostovoy2004}. Tensile strain thus suppresses SE more strongly than DE. As a result, DE becomes relatively more dominant, leading to a lengthening of the helimagnetic ordering with tensile strain. Considering only the direct effect of strain on the $p-d$ hopping integral is thus not sufficient to explain the trend observed experimentally.

It is possible that a secondary electronic effect that is not captured by our DFT calculations can account for the experimental trend of decreasing helimagnetic ordering length with increasing tensile strain. Parameters such as the Hund's energy, $J_H$, or the crystal field splitting may play a more complex role as a function of strain in this correlated material. Biaxial strain alters the crystal field via changes in Fe–O bond lengths and symmetry. However, these effects are already considered in the hopping integrals. Hund’s coupling $J_H$, being largely atomic, is not expected to vary significantly with strain. Furthermore, we find that the preferential ordering length is largely unaffected by the $J_H/U$ ratio (with $U = 3$ eV; see SM Fig. S4).

To account for both the direction and magnitude of the helimagnetic ordering length trend with strain, we consider chemical modifications to \sfo{}. Despite the formal Fe$^{4+}$ valence, strong Fe--O covalency places  SrFeO$_3$ in the negative charge-transfer regime. The DFT occupation matrix yields an Fe $3d$ filling close to $d^6$ (see SM Tab.\,S1), reflecting significant Fe $3d$--O $2p$ hybridization rather than a purely atomic valence. Accordingly, O $2p$ states dominate at the Fermi level (see SM Fig.\,S5), consistent with a covalent $d^5\underline{L}$ description. Our resonant inelastic x-ray scattering (RIXS) data and multiplet calculations, shown in SM Fig. S7, confirm this picture.
Due to this electronic configuration, the formation of oxygen vacancies mainly removes the ligand hole rather than destabilizing the Fe sublattice.
This reduces the energy cost of vacancy formation and makes oxygen non-stoichiometry readily accommodated in SrFeO$_3$ \cite{Ford2020}. 
Tensile strain then further promotes vacancy formation -- not only because vacancies relax the lattice misfit, but also because the local lattice expansion associated with vacancy formation partially offsets the out-of-plane contraction imposed by in-plane expansion~\cite{Lebedev2004,Mayeshiba2017,Aschauer2013}. On the other hand, compressive strain has a much weaker effect on the formation energy of oxygen vacancies \cite{Aschauer2013}.
In this work, we do not model oxygen vacancies explicitly; instead, we capture the electronic effect of oxygen deficiency by tuning the total number of electrons, thereby simulating the associated carrier doping \cite{Takahashi2017}.

Our DFT calculations shown in Fig. \ref{fig:strain+Ov}c,d (also see SM Fig. S6) show that in SrFeO$_{3-x}$ ($0 \leq x \leq 0.1$) the magnetic ordering length decreases systematically with increasing oxygen vacancy concentration $x$. To investigate the origin of this effect, we computed the overlap integral $S$ and the center of mass (COM) for Fe $e_g$ and O 2$p$ states in oxygen-deficient bulk SrFeO$_{3-x}$ (see SM Tab. S2). In the occupied states, $S$ increases for both spin channels (e.g., $S^\uparrow$: 1.373~$\rightarrow$~1.410), and the COM of both orbitals shifts slightly downward, reflecting enhanced Fe($e_g$)–O(2$p$) $p$–$d$ hybridization and modest stabilization of the occupied states. Changes in the unoccupied states are minimal. The enhanced Fe–O hybridization at $x=0.1$ indicates that electron doping associated with oxygen vacancies primarily redistributes electronic spectral weight within the occupied states. This enhanced $p$–$d$ hybridization promotes superexchange interactions over double exchange interactions, consistent with the Mostovoy model \cite{Mostovoy2005}, thereby favoring shorter helimagnetic ordering as the oxygen vacancy concentration increases.

Considering this evidence, the experimental trend of the helimagnetic ordering length can be reproduced by assuming an increasing effective density of electrons from compressive (\lao{}) to tensile (\sto{}) strain conditions, as shown in Fig. \ref{fig:strain+Ov}b. This suggests that the strain-dependent redistribution of vacancy-induced charge carriers, rather than strain alone, plays a central role in determining the helical period.

\section{Discussion and conclusion}

We find that the [111] helimagnetic ordering in \sfo{} is sensitive to biaxial strain, tilting slightly off the [111] vector in the compressed direction, and exhibiting a real-space contraction on the order of 10\% due to an expansion of the [111] unit cell body diagonal on the order of only 1\%. The sign of the strain trend is counter to that observed in B20 or multiferroic helimagnets and inconsistent with theory from both model Hamiltonian and first principles calculations. Instead, we propose that the extreme strain effect is chemically-mediated.

Tensile strain is known to promote the formation of oxygen vacancies \cite{Aschauer2013}, consistent with, and reciprocal to, the established concept of chemical expansion, whereby oxygen vacancy formation produces a positive chemical pressure, in this case further driven by the energetically-favorable reduction of Fe$^{4+}$ toward Fe$^{3+}$ \cite{Chen2013b}. Our DFT calculations show that oxygen loss as low as 3\% is enough to reproduce both the magnitude and direction of the strain effect on the helimagnetic ordering length.

We briefly compare our findings to previous studies on chemical modifications of \sfo{}. 1) Beyond the oxygen off-stoichiometries of a few percent considered here, strontium iron oxides are also stabilized in various oxygen ordered phases. Sr$_8$Fe$_8$O$_{23}$ (SrFeO$_{2.87}$) exhibits helimagnetism with an ordering length of 11 Å \cite{Reehuis2012}. Although the departure from cubic symmetry introduces distinct iron sites and magnetic anisotropy, it is interesting that the loss of oxygen also leads to a shorter helix. Highly oxygen deficient, but not oxygen-ordered, \sfox{} with $x = 2.68$ is not observed to be helimagnetic but may have magnetic sublayers with different exchange interactions \cite{Palakkal2021}. 2) The compound La$_{1/3}$Sr$_{2/3}$FeO$_3$ exhibits a commensurate helimagnetic order with an ordering length of 13.4 Å \cite{Okamoto2010,Ma2011}. La-substitution, like oxygen vacancy formation, induces electron doping and a lattice expansion, so the microscopic origin may be similar. We note, however, that the chemical modification required is an order of magnitude larger with La-substitution (30\%) than with oxygen vacancies to achieve a similar ordering length contraction. 3) We also note that CaFeO$_3$ exhibits incommensurate helimagnetic ordering with a length of 13.5--13.7 Å \cite{Woodward2000,Rogge2019}. CaFeO$_3$ is isovalent with \sfo{} but with a smaller lattice parameter. Crucially, CaFeO$_3$ is insulating while \sfo{} is metallic. In the case of Ca-substitution, the increased localization favors SE over DE and directly results in shorter ordering length. Oxygen stoichiometry may be comparatively less significant. 4) Lastly, a 5\% substitution of cobalt for iron lengthens the helix up to 22 Å \cite{Long2012,Chakraverty2013}. Since SrCoO$_3$ is a ferromagnet, Co-substitution is expected to enhance the DE, which is consistent with our results.

Chemical modifications such as oxygen vacancies may also explain the spread of helimagnetic wavelength values reported in bulk \sfo{}, shown as the gray markers in Fig. \ref{rsxs}c), as well as differences between bulk and thin film samples, and between thin films prepared by different means. Thin films may be more susceptible to oxygen loss than bulk samples, owing to their higher surface area to volume ratio, and samples synthesized and processed in different conditions may not be chemically identical. We note, however, that within the series of samples grown by MBE and annealed in oxygen plasma, and within the series of samples grown by PLD and annealed in ozone plasma, the trend of ordering length with strain is robust.

Despite the extreme strain effect on the spiral structure itself, the transition temperature from paramagnetic to helimagnetic remains at the bulk value of 120 K, independent of strain. This is also in contrast to conventional, DMI-based, helimagnets, where biaxial strain is reported to stabilize the helimagnetic phase towards higher temperatures, and again, highlights the fundamentally different physics at play in \sfo{}.

Our finding that the helimagnetic ordering length can be tuned by biaxial strain has direct implications for future \sfo{}-based devices. Since topologically protected skyrmion-like magnetic structures have a size governed by the underlying helimagnetic order, the ability to engineer their size and distribution is valuable for the development of high density magnetic memories with inherent topological protection. This finding is also crucial for engineering high-energy magnonic devices, since the magnon energy is inversely proportional to the helimagnetic spiral length.

Finally, our observations suggest a possible reconfigurable strategy to manipulate helimagnetic spin textures based on the high mobility of oxygen in \sfox{}. Indeed, \sfox{} compounds are widely studied for their superior oxygen-ion transport properties and potential application in neuromorphic hardware \cite{Nallagatla2019, Ge2019}. In particular, electric field is found to drive a phase transition between the perovskite phase and the brownmillerite \sfobm{} that can be reversed by oxygen annealing \cite{Nallagatla2019}. A similar electrostatic approach could be employed here to fine-tune the helimagnetic ordering \textit{in situ}, offering an alternative to successful magnetoelectric-based tuning e.g. in BiFeO$_3$ \cite{Rovillain2010}.\\

In conclusion, we report that the magnetic ground state of \sfo{} thin films is analogous to prior reports from bulk samples but that biaxial tensile strain significantly shortens the helimagnetic ordering length. We attribute these changes to robust strain-induced modifications in stoichiometry leading to enhanced superexchange interactions. The effect is promising as an engineering strategy towards functional helimagnetic devices for spintronic, magnonic and quantum information applications. Finally, the strain control may be general to other chemically-tunable helimagnets and may provide a platform to understand the origins of this complex ordering in a broad class of quantum materials.

\section{Methods}
\subsubsection*{Sample preparation -- PLD}
Thin film samples were grown by pulsed laser deposition by a KrF excimer laser of 248 nm wavelength operating at 3 Hz. The laser fluence was calculated to be 0.97 J cm$^{-2}$. The target was a pressed powder of \sfobm{}. Samples were grown in 50 mTorr of flowing oxygen at 620 $^\circ$C. 2 nm \sto{} capping layers were grown in-situ at the same conditions as, and immediately following, \sfobm{}. A single crystal of \sto{} served as the target. Warm up and cool down were ramps of 100 $^\circ$C per minute and were carried out in the same oxygen partial pressure as the growth. Ex-situ ozone anneal was performed at 200 - 260 $^\circ$C for 0.5 to 3 hours in a UV-assisted ozone plasma generator.

\subsubsection*{Sample preparation -- MBE}
Thin films were grown via molecular beam epitaxy in an ultra high vacuum chamber with base pressure equal to $3\times10^{-7}$ mTorr. During the deposition, we maintained the substrate temperature at 650 $^o$C in a $4.5\times10^{-3}$ mTorr oxygen background pressure. To completely oxidize the samples, we annealed them up to 600 $^o$C in a 0.01 mTorr oxygen plasma atmosphere and then cooled down slowly under the same conditions.

\subsubsection*{In-house x-ray diffraction}
$\theta$-2$\theta$ measurements were acquired on a PANalytical X'Pert 2 Pro, a Bruker D8 Discover, or a Rigaku Smartlab diffractometer. Reciprocal space maps were acquired on a PANalytical Empyrean.

\subsubsection*{Neutron diffraction}
The samples were measured in the [hhl] scattering plane with the (hhh) direction positioned at low scattering angle ($\theta$ $\approx$ 20$^\circ$) to access and optimize the magnetic satellites with respect to the neutron flux. Measurements were performed at the WISH diffractometer of the ISIS Neutron Source \cite{Chapon2011}.

\subsubsection*{Resonant soft x-ray scattering}
The temperature-dependent $\theta$-2$\theta$ scans were performed using an in-vacuum four circle goniometer at the Resonant Elastic and Inelastic X-ray Scattering (REIXS) beamline of the Canadian Light Source (CLS) \cite{Hawthorn2011}. The samples were mounted on a wedge of 54.7$^\circ$ and rotated around the surface normal by 45$^\circ$ to orient the [111] direction in the horizontal diffraction plane. The scattered x-rays were detected using a silicon drift detector. The ordering length is obtained by fitting the peaks with Gaussian lineshapes and extracting the center.

\subsubsection*{Density functional theory}
We perform first-principles density functional calculations 
within the  Perdew–Burke–Ernzerhof exchange–correlation functional revised for solid (PBEsol) \cite{PBEsol:2008} as implemented in the Vienna {\it Ab initio} Simulation Package ({\sc vasp}) with the projector augmented wave (PAW) method \cite{kresse_ab_1993, kresse_efficiency_1996, kresse_efficient_1996, perdew_generalized_1996, blochl_projector_1994, kresse_ultrasoft_1999}.  The core and valence electrons are treated using the following  electronic configurations:
4$s^{2}$4p$^6$5s$^2$ (Sr),
3d$^7$4s$^1$ (Fe), 
2s$^2$2p$^4$ (O), and a 600~eV plane wave cutoff.
Calculations were carried out using the experimental lattice parameters, while internal distortions were neglected since none are observed experimentally.
The Brillouin zone was sampled with 
a 12$\times$12$\times$12 Gamma-centered k-point mesh.
For the noncollinear spin-order calculations, the wave function is expressed as a spinor, and a generalized Bloch boundary condition is applied~\cite{Sandratskii}. To stabilize the helical spin configuration, the on-site Coulomb interaction $U$ and Hund’s exchange interaction $J_H$ are included~\cite{Li2012a}. In this study, we use 
$U=3.0$\,eV and $J_H=1.0$\,eV. 
The electronic effect of oxygen deficiency is simulated using the background-charge approach, in which the total number of electrons is tuned, and charge neutrality is enforced by a uniform compensating background.

\subsubsection*{Multiplet calculations}
Hybridized ligand field multiplet calculations were carried out following our previous work, utilizing a reduced metal cluster with two ligands along the X and Y directions \cite{Lomeli2025}. Parameters were set based on calculations of other octahedral iron oxide compounds \cite{Ramachandran2025} and are reported in the SM Section 8.

\begin{acknowledgments}
The work at SLAC/Stanford was supported by the US Department of Energy, Office of Basic Energy Sciences, Division of Materials Sciences and Engineering, under contract no. DE-AC02-76SF00515. SJM was supported by the Army Research Office, Grant No. W911NF-15-1-0133.
LB thanks São Paulo Research Foundation - FAPESP for the financial support (Grant 2018/25718-5). Part of the research described in this paper was performed at the Canadian Light Source, a national research facility of the University of Saskatchewan, which is supported by the Canada Foundation for Innovation (CFI), the Natural Sciences and Engineering Research Council (NSERC), the Canadian Institutes of Health Research (CIHR), the Government of Saskatchewan, and the University of Saskatchewan. Part of this work was performed at the Stanford Nano Shared Facilities (SNSF) RRID:SCR\_023230, supported by the National Science Foundation under award ECCS-2026822.
Computational work at Northwestern was supported by the Department of Energy, Office of Science FWP-101256 as part of the Enabling Science for Transformative Energy-Efficient Microelectronics (ESTEEM) Center.
Multiplet calculations were performed on the Sherlock cluster at Stanford University and on resources of the National Energy Research Scientific Computing Center (NERSC), a Department of Energy Office of Science User Facility, using NERSC award BES-ERCAP0027203.
\end{acknowledgments}

\bibliography{PAPERS-SFO,puggioni}%
\end{document}